\documentclass[proceedings, preprint]{rmaa}
\SetYear{2011}
\SetConfTitle{Astronomical Site Testing Data in Chile}

\title{Observing Conditions for Submillimeter Astronomy} 

\author{
  Simon J. E. Radford,\altaffilmark{1} 
}

\altaffiltext{1}{Caltech Submillimeter Observatory,
California Institute of Technology,
  111 Nowelo Street, Hilo, HI 96720, USA 
  (sradford@caltech.edu).}

\shortauthor{Radford}
\shorttitle{Submillimeter Astronomy}

\listofauthors{S. J. E. Radford}
\indexauthor{Radford, S. J. E.}

\abstract{%
Consistently superb observing conditions are crucial for achieving
the scientific objectives of a telescope. 
Submillimeter astronomy is possible only at 
a few exceptionally dry sites, notably
Mauna Kea, the Antarctic plateau, and the 
Chajnantor region in the high Andes
east of San Pedro de Atacama in northern Chile.
Long term measurements of 
225\,GHz and 350\,$\mu$m atmospheric transparency
demonstrate all three locations enjoy significant 
periods of excellent observing conditions.
Conditions on the Chajnantor plateau and at the South Pole
are better more often than on Mauna Kea.
Conditions are better during winter and at night.
Near the summit of Cerro Chajnantor,
conditions are better than on the Chajnantor plateau.
}

\resumen{
\it{%
Condiciones de observaci\'on consistente excelente son cruciales 
para lograr
los objetivos cientficos de un telescopio.
Astronom\'ia submilim\'etrica es posible solamente en
algunos sitios excepcionalmente seco, notablemente
Mauna Kea, la plateau de Ant\'artida, y 
la regi\'on de Chajnantor en los Andes
al este de San Pedro de Atacama en el norte de Chile.
Mediciones \'a largo duraci\'on de la 
transparencia atmosf\'erica
\'a 225\,GHz y 350\,$\mu$m 
demuestran las tres localidades cuentan con suficiente
perodos de excelentes condiciones de observaci\'on.
Las condiciones en el Llano de Chajnantor y en el Polo Sur
son mejores con m\'as frecuencia que en Mauna Kea.
Las condiciones son mejores durante el invierno y por la noche.
Cerca de la cima del Cerro Chajnantor,
las condiciones son mejores que en el Llano de Chajnantor.
}
}

\addkeyword{Atmospheric effects}
\addkeyword{Site testing}
\addkeyword{Submillimeter: general}

\begin{document}
% Typeset article header
\maketitle

\section{Introduction}
\label{sec:intro}

Star formation occurs deep within in dense interstellar molecular clouds permeated 
with obscuring dust.
Heated by the optical and UV radiation emerging from nascent stars,
the dust reradiates at longer, submillimeter wavelengths.
Submillimeter observations provide, therefore, key information for understanding the 
details of star formation and for tracing its progress throughout the history 
of the universe. 
In addition to the continuum, there is a rich line spectrum of 
molecular rotational transitions and atomic fine structure lines
in the submillimeter, which
provide essential data for understanding the cycle of material from the 
interstellar medium through stars and back again.

\begin{figure}[!t]\centering
  \vspace{0pt}
  \includegraphics[viewport=55 60 555 710,clip,
    angle=-90,width=0.8\columnwidth]
    {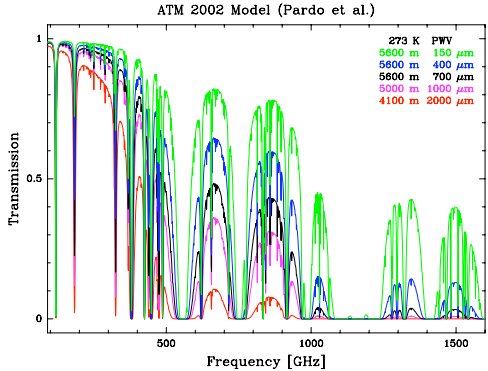}         % color figure
%   {atm-bw}      % black and white version
  \caption{
    Model atmospheric transmission spectra 
    at submillimeter wavelengths for
    different amounts of precipitable water vapor (PWV) 
    in temperate conditions ($0^\circ$\,C) on 
    high altitude (4000--5600\,m) mountains. 
    The relationship between PWV and transmission is not applicable 
    under very cold, i. e., Antarctic conditions.
    Calculated with the ATM model (Pardo et al.\ 2001).
  }
  \label{fig:atm}
\end{figure}

Earth's atmosphere poses an serious impediment to 
submillimeter observations, however,
because water vapor and other molecules absorb strongly at these wavelengths. 
Atmospheric transmission declines dramatically at high frequencies, especially
under wetter conditions
(Fig.\ \ref{fig:atm}).
Observations are possible only through discrete windows
delineated 
by pressure broadened transitions. 

Although spacecraft such as {\it Herschel} 
are not limited by the atmosphere and can provide 
superlative science return, mission costs are high and the practical 
telescope size is limited.
Telescopes on the ground, on the other hand, 
can be much bigger, are
less expensive, and enjoy easier access
but require consistently superb observing
conditions to achieve their scientific objectives. 
It is imperative, therefore, to 
thoroughly characterize a 
potential telescope site 
before embarking on construction.

\section{Sites}
\label{sec:sites}

Worldwide, the typical precipitable water vapor  (PVW) is about
25\,mm (Seidel 2002), high enough to preclude submillimeter observations.
Because of the uneven distribution of
tropospheric water vapor, however, 
a few exceptionally dry locations are suitable for observatories. 
Water vapor is concentrated in the lower troposphere 
with a typical scale height of 1-2\,km, 
much less than the pressure scale height of 7-8\,km.
Deserts, by definition, are arid so
accessible desert mountain summits are good candidates.

Three premiere sites for submillimeter astronomy are Mauna Kea, 
the Antarctic plateau, and the high Andes in northern Chile.
Despite its location in the middle of the Pacific ocean, Mauna Kea 
enjoys good observing conditions because a temperature inversion
frequently traps moisture well below the 4100\,m summit altitude.
Telescopes on Mauna Kea,
including the CSO, the JCMT, and the SMA,
pioneered submillimeter astronomy. 
Even though the air is close to saturated over the ice surface
of the Antarctic plateau, the extreme cold, especially in winter,
means the absolute humidity is very low.
Several telescopes, including the SPT, now
carry out mm and submm observations at the 2835\,m high South Pole.
The higher 
domes C (3230\,m), F (3800\,m), and A (4100\,m)
have been suggested for future observatories
(Spinoglio \& Epchtein 2010).

In northern Chile, the Atacama desert is arguably the driest in the world.
The combined effects of 
a persistent high pressure cell over the eastern Pacific, 
a coastal inversion layer created by
the cold Humbolt current, and
a coastal mountain range
prevent the prevailing westerly winds
from transporting moisture into the region.
To the east, the barrier of the high 
Andean cordillera hinders the flow of
tropical convection
from the Amazon. 
The consequent lack of glaciers, even on the highest peak in the region,
Volc\'an Llullaillaco (6740\,m), is unique for these altitudes
and attests to the aridity
% and the lack of precipitation 
(Messerli et al.\ 1993).
On the 5000\,m high Chajnantor plateau
east of San Pedro de Atacama, 
several small telescopes have made important contributions
and ALMA is nearing completion.
The candidate site for CCAT
is at 5612\,m near the summit of 
Cerro Chajnantor 
overlooking the plateau
(Radford et al.\ 2010).

\begin{figure}[!t]\centering
  \vspace{0pt}
  \includegraphics[viewport=70 55 525 635,clip,
    angle=-90,width=0.8\columnwidth]{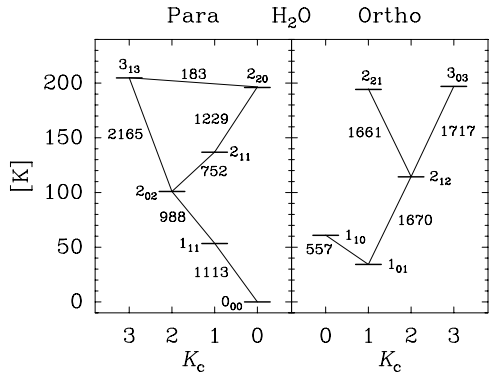}
  \caption{
    Energy levels of the lowest rotational states
    of water with transitions labeled by the frequency in GHz.
    The lines at 557, 752, 988, 1113, and 1229\,GHz
    delineate the principal submillimeter windows.
  }
  \label{fig:water}
\end{figure}

\section{Transparency}
\label{sec:transparency}

Atmospheric transparency is {\it the\/} fundamental 
site characteristic because 
astronomical observations will be futile if cosmic
radiation cannot penetrate the atmosphere.
%Unlike the situation with atmospheric stability, 
%no (obvious)
No countermeasures are available to
improve the intrinsic transparency of 
the atmosphere over a site.

Atmospheric absorption imposes a twofold penalty.
``The degradation in [system] noise temperature due to {\it absorptive 
elements\/} results from two effects: reduction in signal $\ldots$ 
by $\ldots$ absorption and introduction of noise by re-radiation$\ldots$'' 
(Penzias \& Burrus 1973). 
This is true for both coherent (heterodyne) receivers 
and background limited bolometers.
Because integration time is proportional to the square 
of the system noise,
a small fractional improvement in transparency 
can yield a large improvement in observing efficiency.

\section{Atmospheric Spectrum}

At millimeter and submillimeter wavelengths,
the atmospheric spectrum is dominated by strong 
resonant absorption lines of the major molecular species, 
notably H$_2$O, O$_2$, and O$_3$
(Fig.\ \ref{fig:atm}).
Also
pseudo-continuum absorption,
which has both wet and dry components,
arises from the far
wings of strong infrared lines 
and
is stronger at higher frequencies.
Contemporary models, such as ATM (Pardo et al.\ 2001),
{\it am\/} (Paine 2011), and 
Moliere (Urban et al. 2004),
have been verified against 
spectroscopic measurements
(Matsushita et al.\ 1999, Paine et al.\ 2000, Pardo et al.\ 2005).
These models are parameterized by
the barometric
pressure, equivalent to the site
altitude, and the vertical profiles of
air temperature and water vapor. 
Because the shape of the vertical profiles usually can be assumed, 
values of the surface temperature and the total 
PVW are sufficient.

It is worth noting the principal submillimeter windows are delineated
by rotational transitions  
between low energy states of water vapor
(Fig.\ \ref{fig:water}).
For a given atmospheric water vapor content, then,
these states will be more populated, 
and the lines will be stronger, 
if the air temperature is very low.
Detailed model calculations bear this out.
At 225\,K (Antarctic winter),
the ratio of optical depth to PVW 
is about twice as high as at 275\,K (temperate locations).

\begin{figure}[!t]\centering
  \vspace{0pt}
  \includegraphics[viewport=145 75 550 690,clip,
   angle=-90,width=0.8\columnwidth]{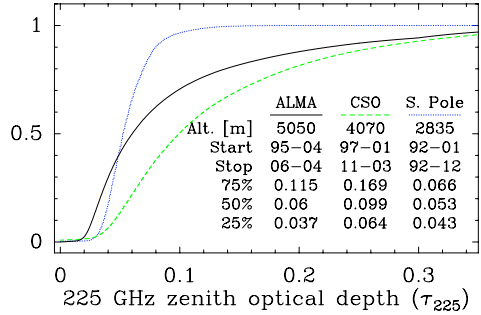}
  \caption{
    Cumulative distributions of 225\,GHz zenith optical depth measured
    with functionally identical narrow band heterodyne tipping radiometers 
    on the Chajnantor plateau (ALMA),
    on the CSO on Mauna Kea,
    and at the South Pole
    (Chamberlin \& Bally 1994; Radford \& Chamberlin 2000). 
    At the CSO,
    $\tau_{225} < 0.06$ is considered suitable for submillimeter 
    observations.
  }
  \label{fig:t225}
\end{figure}

\section {Tipping radiometers}

Microwave sounding of the atmosphere was pioneered 
by Dicke et al.\ (1946),
who inferred the atmospheric absorption 
from measurements of brightness at different zenith angles.
This tipping radiometer technique 
remains standard for 
monitoring atmospheric transparency.
Continuum radiometers
measure the total atmospheric
absoprtion, including all components, directly
at a wavelength of astronomical interest.

\subsection{225\,GHz measurements}

Functionally identical 225\,GHz heterodyne radiometers,
initially developed by NRAO for 
characterizing potential mmA sites,
have been deployed to 
Mauna Kea, to the South Pole, and to the Chajnantor plateau
(Fig.\ \ref{fig:t225}).
All three sites enjoy periods of excellent transparency 
suitable for submillimeter observations,
$\tau_{225}<0.06$.
Mauna Kea, however, experiences these conditions
about half as often as the other two sites.
Poor conditions 
are very rare at the South Pole.
During the best conditions,
the 225\,GHz transparency at the Chajnantor plateau is better than 
at the South Pole.

\begin{figure}[!t]\centering
  \vspace{0pt}
  \includegraphics[viewport=145 75 550 690,clip,
    angle=-90,width=0.8\columnwidth]{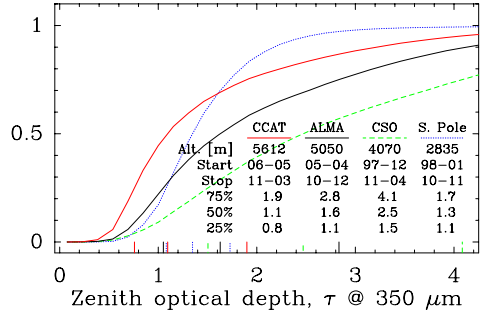}
  \caption{
    Cumulative distributions of 350\,$\mu$m zenith optical depth measured
    with functionally identical broad band tipping radiometers 
    near the summit of Cerro Chajantor (CCAT),
    on the Chajnantor plateau (ALMA), 
    at the CSO on Mauna Kea,
    and at the South Pole. 
    Measurements adjusted for radome transparency
    (Calisse 2004).
    }
  \label{fig:t350}
\end{figure}

\subsection{350\,$\mu$m measurements}

Because of the uncertainties inherent in extrapolating from
225\,GHz measurements to higher frequencies,
broadband tippers were deployed to
directly measure the 350\,$\mu$m atmospheric transparency.
These instruments were developed jointly by NRAO and Carnegie Mellon University.
The distributions
of the 350\,$\mu$m measurements (Fig.\ \ref{fig:t350})
and the 225\,GHz measurements (Fig.\ \ref{fig:t225})
are quite similar.
The first quartile 350\,$\mu$m transparencies at the Chajnantor
plateau  
and the South Pole are roughly equal and noticeably better than
at Mauna Kea. 

\begin{figure}[!t]\centering
  \vspace{0pt}
  \includegraphics[viewport=50 70 535 710,clip,
   width=0.8\columnwidth]{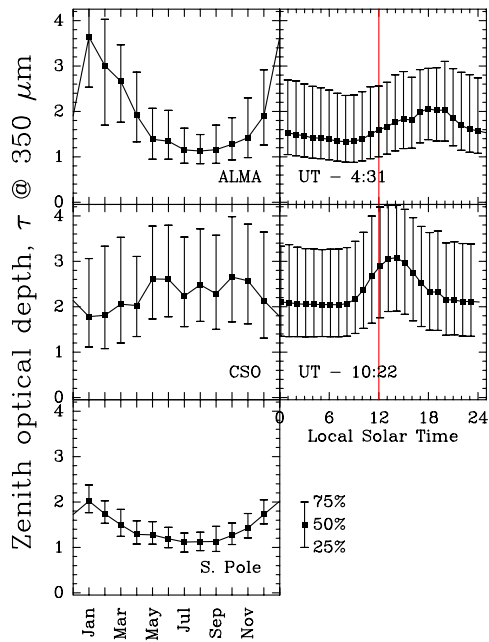}
  \caption{
    Variations of the 350\,$\mu$m zenith 
    optical depth measured
    and on the Chajnantor plateau (ALMA),
    at the CSO on Mauna Kea,
    and at the South Pole. 
    The error bars show the 
    quartiles for each month or hour.
   }
  \label{fig:t350-var}
\end{figure}

Seasonal variations are significant 
(Fig.\ \ref{fig:t350-var}).
At all three sites, 
the transparency is better during the winter 
than the summer. 
At the South Pole, conditions are
remarkably consistent from year to year.
The seasonal pattern for Mauna Kea, although evident,
is overshadowed by year-to-year variations.
In the Chajnantor region, conditions are consistently good
from April through December but 
deterioriate during the summer months
when a shift in atmospheric 
circulation patterns draws moist air 
over the Andes from the Amazon basin.
There is considerable year-to-year variation in the severity 
of this summer season.
Winter conditions in the Chjanantor region
have less interannual
variability than at Mauna Kea. 

Both the Chajnantor region and Mauna Kea experience diurnal
transparency variations, with better conditions
at night.
These diurnal transparency variations 
lag behind the solar cycle.
The best conditions occur around sunrise.
At Chajnantor, the diurnal variations are weaker 
during the winter than during the summer.

\section{Cerro Chajnantor}

In the Chajnantor region, several peaks rise more than 500\,m
above the plateau.
Radiosondes launched from the plateau 
measured the water vapor density profiles and 
have determined the typical scale height is
about 1.1\,km.
The flights also revealed inversion layers frequently occur, 
especially at night,
that trap moisture below the below the height of these peaks
(Giovanelli et al. 2001).

\begin{figure}[!t]\centering
  \vspace{0pt}
  \includegraphics[viewport=70 75 550 690,clip,
    angle=-90,width=0.8\columnwidth]{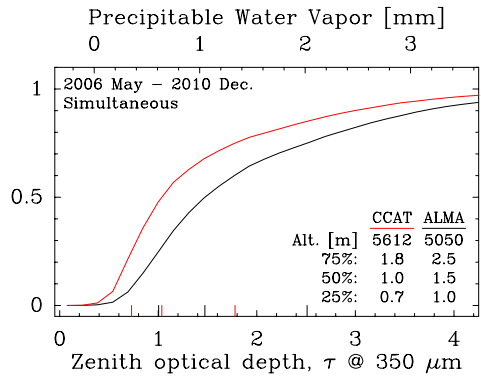}
  \caption{
    Cumulative distributions of 350\,$\mu$m zenith optical depth measured
    simultaneously  
    near the summit of Cerro Chajantor (CCAT)
    and on the Chajnantor plateau (ALMA). 
    This PVW scale is only appropriate for the Chajnantor region.
  }
  \label{fig:cchaj}
\end{figure}

CCAT is a proposed 25\,m diameter telescope for 
submillimeter astronomy (Radford et al.\ 2010).
The candidate CCAT site is a level bench 5612\,m high
near the summit of Cerro Chajnantor accessed by 
a road constructed by the University of 
Tokyo.
With two tipping radiometers, one at the summit and the 
other on the plateau, the 350\,$\mu$m transparency was measured
simultaneously.
Prior to deployment to Cerro Chajnantor, the instruments were 
were operated side by side 
on the plateau where they reported identical results.
Observing conditions for CCAT are
significantly better
than on the plateau
(Fig.\ \ref{fig:cchaj}),
offering more opportunities 
for observations at short wavelengths.

The 350\,$\mu$m optical depth measurements
on the Chajnantor plateau
are well correlated 
with PWV measurements from a 183\,GHz spectrometer
on the APEX telescope.
Linear regression then
indicates
PWV${} = 0.935\, (\tau_{350} - 0.35)$.
   This relation is only appropriate in the environs of Chajnantor.

\adjustfinalcols

\acknowledgements

It is a pleasure to thank the
many colleagues who have contributed to these measurements. 
Jeff Peterson initiated development of the 350\,$\mu$m tippers.
CCAT site evaluation is carried out in the Parque Astron\'omico Atacama 
in northern Chile under the auspices of the Programa de Astronom\'ia, 
a program of the Comisi\'on Nacional de Investigaci\'on Cient\'ifica y 
Tecnol\'ogica de Chile (CONICYT).
CCAT site evaluation received partial support from the
National Science Foundation (AST-0431503).
The NRAO is a facility of the NSF operated under a cooperative
agreement by Associated Universities, Inc.
The Caltech Submillimeter Observatory (CSO) is operated by the 
California Institute of Technology with funding from the NSF
(AST-0838261).

\end{document}